\documentclass[twocolumn,twoside]{opticajnl}
\journal{opticajournal} % use for journal or Optica Open submissions

\usepackage{lineno}
\DeclareUnicodeCharacter{2212}{-}

\title{Heralded pure single-photon sources using nanophotonic waveguides with quadratic and cubic nonlinearities}

\author[1,*]{Mahmoud Almassri}
\author[1]{Mohammed F. Saleh}

\affil[1]{MNO group,Institute of Photonics and Quantum Sciences, Heriot-Watt
University, EH14 4AS Edinburgh, UK}
\affil[*]{mha2003@hw.ac.uk}

\begin{abstract}

This paper presents, to our knowledge, a new approach in developing integrated pure heralded single-photon sources based on the interplay between the spontaneous four-wave mixing  and sum-frequency generation parametric processes. We introduce a comprehensive quantum model to exploit this interplay in AlGaAs and LiNbO$_3$ nanophotonic waveguides. The developed model is used to assess the performance of the sources based on the photon-pair generation and the associated spectral purity. We find that this approach can remarkably improve the spectral purity of low-pure generated photon pairs, relaxing the restrictions on the structure design and the used pump wavelength. In addition, it overcomes the current hurdles in implementing on-chip photon detectors operating at room temperature, paving the way for advanced applications in integrated quantum photonics and information processing.\\
\end{abstract}

\setboolean{displaycopyright}{False} % Do not include copyright or licensing information in submission.
\begin{document}

\maketitle

\section{Introduction}
Quantum photonics has emerged as a transformative field for exploiting the control of light at the quantum level in pioneering the next-generation technologies \cite{c01,c02}. In the centre of this advancement is a single-photon source (SPS) that is a cornerstone for a variety of applications, ranging from quantum computing to secure communication. The development and optimisation of an on-chip SPS are also critical for achieving compact scalable quantum  systems and   architectures \cite{c03,c04,PhysRevLett.132.133603,PhysRevLett.133.083803}.

Single photons can be deterministically generated  using  quantum dots \cite{c05,c07}. However, practically they are less suitable for waveguide integration, because photons are usually emitted out-of-plane. Also, having multiple identical quantum-dot sources represents a major challenge, due to the difficulty in having dot-to-dot uniformity and inhomogeneous broadening. Alternatively, spontaneous parametric down-conversion (SPDC) or spontaneous four-wave mixing (SFWM) can be utilised in generating probabilistic photon pairs. Single-photon sources can then be engineered  by heralding the existence of one of the two photons. Despite the probabilistic nature of the latter technique that can pose synchronization challenges in quantum networks, it becomes the practical scheme for achieving single photons. Moreover, advances in multiplexing techniques  compensate for the probabilistic nature of photon generation and enhance the system efficiency \cite{c06,c08,c09,Yu:22}.

Recent advances in fabrication techniques have enabled nanophotonic waveguides with enough intensities to excite intrinsic third-order  $\chi^{(3)}$ nonlinearities in non-centrosymmetric materials \cite{apiratikul_enhanced_2014,c014}. Also, simultaneous exploitation of second-order nonlinear $\chi^{(2)}$ interactions in these structures has been facilitated via altering the pattern of $\chi^{(2)}$ to achieve quasi-phase matching \cite{c013,thomas_quasi_2013,morais_directionally_2017,zhang_high_performance_2021}. 

\begin{figure} 
    \centering \includegraphics[width=0.54\columnwidth]{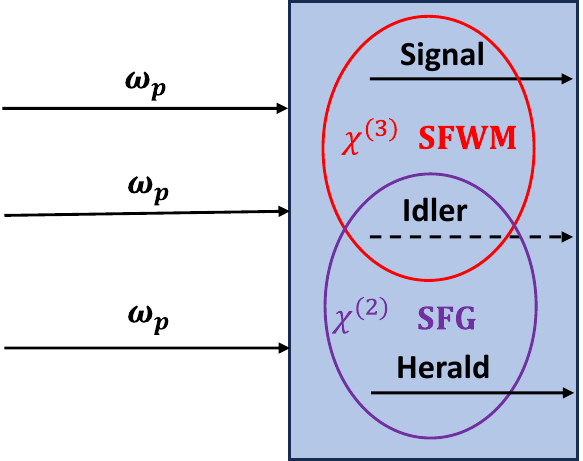} 
    \caption{A schematic diagram explaining the concept of the proposed device, and SFWM and SFG interaction.}
    \label{fig:0} 
\end{figure}

In this work, we propose a new integrated heralded SPS that exploits these recent advancements. The concept is based on the interplay between the SFWM and  sum frequency generation (SFG) parametric processes. Two pump photons are annihilated to generate two sidebands signal and idler photons via the $\chi^{(3)} $SFWM process, simultaneously the latter photon is upconverted using another pump photon in a $\chi^{(2)} $SFG process, as depicted in Fig. \ref{fig:0}.
The up-converted photon is used in heralding the signal photon. The proposed scheme tackles a significant challenge, which is an efficient on-chip detection of the heralding photon. Photon pairs are usually generated in the telecom regime with energies that fall beyond the activation threshold of the on-chip silicon photon detectors, for instance. Moreover, off-chip  superconducting nanowire detectors have a limited  frequency range in addition to the requirement of having cryogenic-temperature operation \cite{c015,c016,c017,c018}.

The paper is organised as follows. In Sec. 2, we develop a quantum model to efficiently investigate the interplay between SFWM and SFG using continuous and pulsed-pump sources.  In Sec. 3, we implement and simulate the developed approach for different nanophotonic waveguides with optimised geometries and modulation period of the $\chi^{(2)}$ coefficient. Finally, our conclusions are summarised in Sec. 4.

\section{Modelling simultaneous $\chi^{(2)}$ and $\chi^{(3)}$ nonlinear interactions}
In this section, we develop a quantum model  to study the interplay between the SFWM and SFG in a waveguide with simultaneous $\chi^{(2)}$ and $\chi^{(3)}$ nonlinearities. The model is based on the Heisenberg equation of motion for deriving a set of coupled differential equations that describe the spatial evolution of the creation and annihilation operators of the signal, idler, and up-converted (heralding) photons along the waveguide. The Hamiltonian is determined by integrating the energy flux flows through a cross-sectional area of the waveguide over a very long quantisation time  $T$ \cite{C1,C2}. The model is developed first for the continuous pump wave then generalised for the pulsed pump case. The self-phase, cross-phase modulations, and periodic alternation of $\chi^{(2)}$ have been taken into account.

\subsection{Continuous-wave (CW) regime }
Assuming the pump is an undepleted  monochromatic classical field with a frequency $\omega_{p}$ and a propagation constant $k_{p}$,  its electric field can be written as

\begin{equation}
\begin{split}
      E_{p}(z,t)=\frac{1}{2}\left[A_{p}\exp{(-j(\omega_{p}t-k_{p}z))}+c.c.\right],
\end{split}
    \label{eqn:pumpelectric}
\end{equation}
where $z$ is the longitudinal direction, $t$ is the time, $A_{p}$ is the wave amplitude, and $c.c.$ stands for the complex conjugate. The quantised electric fields of the signal ($s$), idler ($i$), and heralding ($h$) photons resulting from the SFWM and SFG processes are given by \cite{C1,C3}

\begin{equation}
    \hat{E}_q(z,t)=\hat{E}_q^{+}(z,t)+\hat{E}_q^{-}(z,t),
    \label{eqn:Eoperator}
\end{equation}

\begin{equation}
    \hat{E}_q^{+}(z,t)= \sum_{m}\sqrt{\frac{\hbar \omega_{m}}{2 \epsilon_{0} c T n_{m}}} \hat{a}\left(z, \omega_{m}\right) \exp{(-j \omega_{m} t)},
    \label{eqn:Eoperator}
\end{equation}
where $\hat{E}_q^{-}(z,t)=\left[\hat{E}_q^{+}(z,t)\right]^\dagger$, $\epsilon_{0}$ is the free space permittivity, $\hat{a}\left(z, \omega_{m}\right)$ is the  annihilation operator of the mode $m$ with frequency $\omega_{m}$ and refractive index $n_{m}$, $\hbar$ is the reduced Planck's constant, and $q=s,i,h$.

In the Heisenberg picture, the space evolution of the creation or annihilation operators is defined as \cite{C4}

\begin{equation}
    -j\hbar \frac{\partial \hat{a}(z,\omega_{m})}{\partial z}=[\hat{a}(z,\omega_{m}),\hat{G}],
    \label{eqn:motionequation}
\end{equation}
where $\hat{G}$ is the momentum operator  calculated by integrating the momentum flux $\hat{g}(z,t)=\hat{D}^{-}(z,t)\hat{E}^{+}(z,t)+H. c.$ over the quantisation time period $T$ with  $\hat{D}(z,t)=\epsilon _0 n\hat{E}(z,t)+\hat{P}_{nl}(z,t)$  the electric-displacement field operator, $\hat{P}_{nl}(z,t)$  the nonlinear  polarisation operator, and $H.c.$ the Hermitian conjugate.

\subsubsection{Linear propagation}
% \noindent \textbf{Propagation in a linear medium} \:
Using Eq. (\ref{eqn:Eoperator}) and setting $\hat{P}_{nl}=0$,  the momentum flux operator for the linear part of propagation can be calculated as

\begin{equation}
\begin{split}
        \hat{g}_{L}(z,t)=&\sum_{m,n}  \frac{\hbar\epsilon}{2 \epsilon_{0} c T } \sqrt{\frac{ \omega_{m}\omega_{n}}{ n_{m} n_{n}}} \hat{a}\left(z, \omega_{m}\right) \hat{a}^{\dagger}\left(z, \omega_{n}\right)\\
        &\times \exp{\left(j(\omega_{n}- \omega_{m})t\right)}+H.c.   
\end{split}
    \label{eqn:LinearFlux}
\end{equation}

\noindent Subsequently the momentum operator $\hat{G}_{L}$ is, 
\begin{equation}
    \hat{G}_{L}(z)=\int_{0}^{T} \hat{g}_{L}( z,t) dt   =\sum_{m}\hbar k_{m} \hat{a}^{\dagger}\left(z, \omega_{m}\right) \hat{a}\left(z, \omega_{m}\right),
    \label{eqn:LinearMomentum}
\end{equation}
using the identity $\int_{0}^{T}e^{j\Delta \omega t} dt=2\pi\delta(\Delta \omega)$. Hence the linear propagation of the $q$-th mode is governed by,
\begin{equation}
    \frac{\partial \hat{a}^{\text{L}}_{q}}{\partial z}= jk_{q}\hat{a}_{q},
    \label{eqn:linearPart}
\end{equation}
with $k_{q}=n_{q}\omega_{q}/c$ and $\hat{a}(z,\omega_{q})=\hat{a}_{q}$ for brevity.
  
\subsubsection{Spontaneous four-wave mixing}
To determine the contribution of the nonlinear SFWM to the evolution of the operators,  we  start by writing the nonlinear polarisation operator of this process
\begin{equation}
   \hat{P}_{\text{SFWM}}=3\epsilon_{0}\chi^{(3)} E^{+}_{p}E^{+}_{p} \hat{E}^{-}_{s}+H.c.
    \label{eqn:SFWMpolarisation}
\end{equation}
The momentum flux operator of the SFWM process can be hence calculated  as

\begin{equation}
   \hat{g}_{\text{SFWM}}=3\epsilon_{0} \chi^{(3)} E^{+}_{p}E^{+}_{p} \hat{E}^{-}_{s}\hat{E}^{-}_{i}+H.c.,
    \label{eqn:SFWMflux}
\end{equation}
and the momentum operator is  therefore
\begin{equation}
\hat{G}_{\text{SFWM}}=\frac{3 \hbar \chi^{(3)} A_p^2 e^{j 2 \kappa_p z}}{16 c} \sum_s \sqrt{\frac{\omega_s \omega_i}{n_s n_i }}  \hat{a}_s^{\dagger} \hat{a}_i^{\dagger}+H.c.,
    \label{eqn:SFWMmomentum}
\end{equation} 
 with the pump self-phase modulation effect included in the modified propagation constant $\kappa_{p}=k_{p}\left(1+\frac{n_{2} I_{p}}{ n_{p} } \right)$,  $n_{2}$ the nonlinear refractive index, and $I_{p}$ the pump field intensity. The subscripts $s$ and $i$  refer to the signal and idler modes that are linked through the energy conservation $2\omega_{p}=\omega_{s}+\omega_{i}$. As a result the spatial propagation of the annihilation operator of the mode $\omega_{s}$ due to the SFWM is 

\begin{equation}
\frac{\partial \hat{a}_s^{(\text{SFWM})}}{\partial z}=j \frac{n_{2} n_{p} I_{p}  e^{j 2 \kappa_p z}}{ c} \sqrt{\frac{\omega_s \omega_i}{n_s n_i  }}    \; \;\hat{a}_i^{\dagger}.
\end{equation}

% \vspace{0.5 cm}
% \noindent\textbf{Cross-Phase Modulation (XPM)}
%  \vspace{0.2 cm}

\subsubsection{Cross-phase modulation (XPM)}
The contribution of the cross-phase modulation (XPM) effect of the pump wave on the propagation of the signal and idler photons can be determined in a similar approach to the SFWM-contribution.  The nonlinear polarisation operator of this process is
\begin{equation}
   \hat{P}_{\text{XPM}}=6\epsilon_{0}\chi^{(3)} E^{+}_{p}E^{-}_{p} \hat{E}^{-}_{s}+H.c.
    \label{eqn:XPMpolarisation}
\end{equation}
Hence, the momentum operator is defined as 
\begin{equation}
\hat{G}_{\mathrm{XPM}}=\frac{\hbar n_{2}n_{p} I_{p}}{c} \sum_s \frac{\omega_s}{n_s}  \hat{a}_s\hat{a}_s^{\dagger}+H.c.,
\label{eqn:XPMMomentum}
\end{equation}
and  the spatial evolution of the annihilation operator due to XPM can be written as
\begin{equation}
\frac{\partial \hat{a}_s^{(\mathrm{XPM})}}{\partial z}=j \frac{ 2 n_{2}n_{p} I_{p}}{c} \frac{\omega_s}{n_s }  \hat{a}_s.
\label{eqn:XPMannihilation}
\end{equation}

\subsubsection{Sum-frequency generation}  
In this second-order nonlinear process, the idler combines with another pump photon to produce a photon at a frequency equals to $\omega_{h}=\omega_{p}+\omega_{i}$. The associated nonlinear polarisation in this case is
\begin{equation}
    \hat{P}_{\text{SFG}}=\epsilon_{0} \chi^{(2)}E^{+}_{P}\hat{E}^{+}_{i}+H.c.,
    \label{eqn:SFGpolarisation}
\end{equation}
and the momentum-flux operator is  determined, similar to other nonlinear processes, as
\begin{equation*}
    \begin{split}
\hat{g}_{\text{SFG}}=\frac{\epsilon_{0} \chi^{(2)}}{2}\sum_{i,h}&A_{p}\exp{(j\kappa_{p}z)}  \frac{\hbar}{2 \epsilon_{0} c T } \sqrt{\frac{ \omega_{i}\omega_{h}}{ n_{i} n_{h}}}\; \hat{a}_{i} \hat{a}^{\dagger}_{h} \\
&\times \exp{\left(-j( \omega_{i}+\omega_{p}-\omega_{h})t\right)}+H.c.
\end{split}
\end{equation*}
Integrating over the time, the momentum operator is found as 
\begin{equation}
    \begin{split}
\hat{G}_{\text{SFG}}=\frac{\hbar}{4}  \sum_{i}\frac{\chi^{(2)}A_{p}}{c}\sqrt{\frac{ \omega_{i}\omega_{h}}{n_{i}n_{h} }} \: \hat{a}_{i}\hat{a}^{\dagger}_{h}\exp{(j\kappa_{p}z)}
+H.c.,\\
\end{split}
    \label{eqn:SFGmomentum}
\end{equation}
and the spatial evolution of the $h$ mode creation operator can then be calculated using Eq. (\ref{eqn:motionequation})
\begin{equation}
    \frac{\partial \hat{a}^{\dagger}_{h}}{\partial z}= -j\frac{\chi^{(2)}A_{p}}{2c}\sqrt{\frac{ \omega_{i}\omega_{h}}{n_{i}n_{h} }}\hat{a}^{\dagger}_{i}\exp{(-j\kappa_{p}z)} .
    \label{eqn:heraldcreation}
\end{equation}
A similar equation can be obtained for the idler by swapping the indices $i,h$. 

\subsubsection{Total effects}
Combining all the linear, nonlinear SFWM, XPM, and SFG polarisation effects, the following set of three coupled differential equations can be written as 
\begin{equation}
 \begin{split}
     &\frac{\partial \hat{a}_{s}}{\partial z}=j\kappa_{s}\hat{a}_{s}+j\gamma_{s,i}\hat{a}_{i}^{\dagger}e^{2j\kappa_{p}z},\\
     &\frac{\partial \hat{a}_{i}^{\dagger}}{\partial z}=-j\kappa_{i}\hat{a}_{i}^{\dagger}-j\gamma_{s,i}\hat{a}_{s}e^{-2j\kappa_{p}z}-j\gamma_{i,h}\hat{a}_{h}^{\dagger}e^{j\kappa_{p}z},\\
     &   \frac{\partial \hat{a}_{h}^{\dagger}}{\partial z}=-j\kappa_{h}\hat{a}_{h}^{\dagger} -j\gamma_{i,h}\hat{a}_{i}^{\dagger}e^{-j\kappa_{p}z},
     \end{split}
     \label{eqn:Qmodel}
 \end{equation}
where $\gamma_{s,i}=\frac{  n_{p} n_{2} I_{p}}{c} \sqrt{\frac{\omega_{s} \omega_{i}}{n_{s} n_{i}  }}$  , $\gamma_{i,h}=\frac{\chi^{(2)}A_{p}}{2c}\sqrt{\frac{ \omega_{i}\omega_{h}}{n_{i}n_{h} } }$  are the coupling coefficients of the SFWM and SFG processes, respectively, and $\kappa_{q}=k_{q}\left(1+\frac{2n_{2} n_{p} I_{p}}{ n_{q}^{2}  } \right)$ is the modified propagation constant of the  $q-$mode due to the XPM effects. Solving the set of the three coupled equations results in relating  the creation and annihilation operators of each mode $q$ at a position $z=z_{0}$ in terms of their corresponding at the waveguide input. %Subsequently, the expected number of photons of a mode $q$ can be computed using the number operator defined as $\hat{N}(z_{0},w_{q})=\hat{a}_{q}^{\dagger}(z_{0})\hat{a}_{q}(z_{0})$.

\subsubsection{Transfer-matrix method}
Via discretising the waveguide into infinitesimal elements, the set of differential equations can be solved within each element \cite{C2}.  Then, a $3\times3$ transfer-matrix $\mathcal{T}$ can be obtained to link the outcome of each element to its input. By multiplying the transfer matrices of all the elements in a descending order, the output of the last element can be  related to the input of the first element. However, we found that directly applying  this method is very computationally expensive, since  the step-size must be much less than the modulation period of the second-order nonlinear coefficient, and each element has a different transfer matrix. 

To circumvent this challenge, we have introduced the following phase-transformation equations
\begin{equation}
    \begin{split}
        &\hat{a}_{s}=\hat{b}_{s}\exp{\left(j\frac{\Delta k_{1}+\Delta k_{2}+2\kappa_{s}}{2}z\right)},\\         &\hat{a}_{i}^{\dagger}=\hat{b}_{i}^{\dagger}\exp{\left(j\frac{\Delta k_{2}-\Delta k_{1}-2\kappa_{i}}{2}z\right)},\\        &\hat{a}_{h}^{\dagger}=\hat{b}_{h}^{\dagger}\exp{\left(j\frac{3\Delta k_{2}-\Delta k_{1}-\kappa_{h}}{2}z\right)}.\\
    \end{split}
    \label{eqn:SecondPhaseTransformation}
\end{equation}
 to remove the $z$-dependence of the transfer-matrix of each element, making them all identical. Substituting these equations in Eq. (\ref{eqn:Qmodel}), the set of coupled  differential equations can be simplified to be
\begin{equation}
\begin{split}
     &\frac{\partial \hat{b}_{s}}{\partial z}=-j\frac{\Delta k_{1}+\Delta k_{2}}{2}\hat{b}_{s}+j\gamma_{s,i}\hat{b}_{i}^{\dagger},\\
     &\frac{\partial \hat{b}_{i}^{\dagger}}{\partial z}=-j\frac{\Delta k_{2}-\Delta k_{1}}{2}\hat{b}_{i}^{\dagger}-j\gamma_{s,i}\hat{b}_{s}-j\gamma_{i,h}\hat{b}_{h}^{\dagger},\\
     &   \frac{\partial \hat{b}_{h}^{\dagger}}{\partial z}=-j\frac{3\Delta k_{2}-\Delta k_{1}}{2}\hat{b}_{h}^{\dagger} -j\gamma_{i,h}\hat{b}_{i}^{\dagger},
     \end{split}
      \label{eqn:QuamtumModel}
\end{equation}
with $\Delta k_{1}=2\kappa_{p}-\kappa_{i}-\kappa_{s}$ and $\Delta k_{2}=\kappa_{h}-\kappa_{i}-\kappa_{p}$. Solving the new set of differential equations over a small step size $\Delta z$, the output operators at the waveguide exit $z=L$ can be written in terms of their inputs as 
\begin{equation}
    \begin{bmatrix}
\hat{b}_{s} \\
\hat{b}^{\dagger}_{i}  \\
\hat{b}^{\dagger}_{h}
\end{bmatrix}_{z=L}=
\mathcal{T}^{N}  \begin{bmatrix}
\hat{b}_{s} \\
\hat{b}^{\dagger}_{i}  \\
\hat{b}^{\dagger}_{h}
\end{bmatrix}_{z=0},
\label{eqn:matrixequation}
\end{equation}
with
\begin{equation}
\mathcal{T}= \begin{bmatrix}
1-j\phi_1 &f_{s,i} & 0\\
f_{s,i}^{*} & 1-j\phi_2 & f_{i,h}\\
0 &  f_{i,h}& 1-j\phi_3
\end{bmatrix} ,  
\end{equation}
$N=\frac{L}{\Delta z}$  the number of steps, $ f_{s,i}=j\gamma_{s,i}\Delta z$ and $f_{i,h}=-j\gamma_{i,h} \Delta z$,  $\phi_1 = \frac{\left(\Delta k_{1}+\Delta k_{2}\right)\Delta z}{2} $, $\phi_2=\frac{\left(\Delta k_{2}-\Delta k_{1}\right)\Delta z}{2} $, and   $\phi_3=\frac{\left(3\Delta k_{2}-\Delta k_{1}\right)\Delta z}{2} $.

\subsubsection{Expected number of photons}
 The expected number of photons of a mode $q$ at the  waveguide output can be calculated using the number operator defined as,
\begin{equation}
    \langle N_{q}(L)\rangle= \langle\psi| \hat{b}^{\dagger}_{q}(L)\hat{b}_{q}(L)|\psi\rangle,
\end{equation}
where $|\psi\rangle= |0\rangle_{s}|0\rangle_{i}|0\rangle_{h}$ is the vacuum state, and  $ q=s,i,h$. Using Eq. (\ref{eqn:matrixequation}), the signal operator is $\hat{b}_{s}(L)=\textbf{T}_{11}\hat{b}_{s}(0)+\textbf{T}_{12}\hat{b}^{\dagger}_{i}(0)+\textbf{T}_{13}\hat{b}^{\dagger}_{h}(0)$, with $\textbf{T}_{uv}$  the matrix element  in the $u$th row and $v$th column of $\mathcal{T}^{N}$. Similar equations can be obtained for $\hat{b}_{i}(L)$ and $\hat{b}_{h}(L)$. Hence, the expected number of photons of the three interacting photons at  $z=L$ are   $   \langle N_{s}(L)\rangle=\left|\textbf{T}_{12}\right|^{2}+\left|\textbf{T}_{13}\right|^{2}$, $    \langle N_{i}(L)\rangle=\left|\textbf{T}_{21}\right|^{2}$, and $ \langle N_{h}(L)\rangle=\left|\textbf{T}_{31}\right|^{2}$.

\subsection{Pulsed-pump regime}
A pulsed-pump source can be viewed as a superposition of multiple monochromatic pump waves. Therefore, the pump electric field can be written in this case as 
\begin{equation}
    E_{p}=\frac{1}{2}\left[\sum_{p} A_p  \Delta \omega_p \;e^{-j\left(\omega_p t-k_p z\right)} +c.c\right],
    \label{eqn:pulsedPump2}
\end{equation}
with $\Delta \omega_p$  the sampling frequency of the pump pulse in the Fourier domain. The amplitude of the harmonic $\omega_p$ is $ A_p=\frac{A_0 \tau}{\sqrt{2 \pi}} e^{-\frac{\tau^2\left(\omega_p-\omega_{p_0}\right)^2}{2}}$, assuming a Gaussian pump $A_0  e^{-t^2 / 2 \tau^2}$ with an amplitude $A_0$, a central frequency $\omega_{p_{0}}$ and a temporal width $\tau$.

\subsubsection{Spontaneous four-wave mixing}
The momentum flux operator of the SFWM process results from every possible combination of two monochromatic pump components with amplitudes $A_{p_{1}}$ and $A_{p_{2}}$ and  frequencies $\omega_{p_{1}}$ and $\omega_{p_{2}}$ is 
\begin{equation}  
    \begin{split}
\hat{g}_{\text{SFWM}}=\frac{3\hbar \epsilon_{0}\chi^{(3)}}{8 \epsilon_{0} c T}&\sum_{p_{1},p_{2},s,i} \sqrt{\frac{\omega_{s} \omega_{i}}{n_{s} n_{i} }} A_{p_{1}}A_{p_{2}}  (\Delta \omega_p)^{2} \hat{a}_{i}^{\dagger} \hat{a}_{s}^{\dagger}\\
&\;\;\;\;\;\;\;\times e^{j\left(k_{p_{1}}+k_{p_{2}} \right)z}  e^{-j\Delta \omega t}+H.c.,  
    \end{split}
    \label{eqn:PulseSFWMg}
\end{equation} 
with $\Delta \omega =\omega_{p_{1}}+\omega_{p_{2}}-\omega_{s}-\omega_{i}$. Integrating $\hat{g}_{\text{SFWM}}$ to get the momentum operator $\hat{G}_{\text{SFWM}}$, a Dirac delta function $\delta(\omega_{s}+\omega_{i}-\omega_{p_{1}}-\omega_{p_{2}})$ is obtained, which is used to cancel the summation over the second pump pulse, $\sum_{p_{2}} \Delta \omega_p=1$, and forces $\omega_{p_{2}}$ to satisfy the energy conservation condition $ \omega_{p_{2}}=\omega_{s}+\omega_{i}-\omega_{p_{1}}$ with the pair of modes $\omega_{s}$ and $\omega_{i}$.
 
Applying Eq. (\ref{eqn:motionequation}), the spatial propagation of the annihilation operator of the $s$ mode due to SFWM in the pulsed-pump regime is

\begin{equation}
\begin{split}
    \frac{\partial \hat{a}_s^{(\text{SFWM})}}{\partial z}=&j \frac{3\Delta \omega_{s,i}\Delta \omega_p \chi^{(3)}}{4c}
    \sum_{p_{1},i}\sqrt{\frac{\omega_s \omega_i}{n_s n_i  }}\\
    &\times A_{p_{1}} A_{p_{2}}
 e^{j\left(\kappa_{p_{1}}+\kappa_{p_{2}}\right)z} \hat{a}_i^{\dagger},
    \end{split}
\label{eqn:SFWMaspulsed}
\end{equation}
where $\Delta \omega_{s,i}=\sqrt{\Delta \omega_{s}\Delta \omega_{i}}=2\pi/T$, with $\Delta \omega_{s}$ and $\Delta \omega_{i}$  the frequency sampling of the signal and the idler photons.  Also, $k_{p_{1}}$ and $k_{p_{2}}$ are replaced by $\kappa_{p_{1}}$ and $\kappa_{p_{2}}$, respectively, to take into account the self-phase modulation effect of the two pump waves.

\subsubsection{Cross-phase modulation} 
Following the same approach for the CW case, we find that the spatial propagation of the signal and idler operators due to cross-phase modulation from the two pump waves can be described as 

\begin{equation}
\frac{\partial \hat{a}_q^{(\mathrm{XPM})}}{\partial z}=j \frac{ 3\chi^{(3)}}{4c} \frac{\omega_q}{n_q } \left[A_{p_{1}}^{2} +A_{p_{2}}^{2} \right]\hat{a}_q,
\label{eqn:XPMannihilation}
\end{equation}
where $q=s,i$, and the two components of the pump waves  satisfy the energy conservation condition $\omega_{p_{1}}+\omega_{p_{2}}=\omega_{s}+\omega_{i}$. The cross-phase modulation between the two pump waves involved in the process can also be taken into account by modifying $\kappa_{p_{1}}$ and $\kappa_{p_{2}}$ used in Eq. (\ref{eqn:SFWMaspulsed}) as follows
\begin{equation}
\kappa_{p_{u}}=k_{p_{u}}\left[1+\frac{3 \chi^{(3)}}{8 n_{p_{u}}^2 }\left(A_{p_{u}}^2 +2 A_{p_{v}}^2 \right) \right],
\label{eqn:SPMandXPM}
\end{equation}
with $u,v=1,2$ and $u\neq v$.

\subsubsection{Sum-frequency generation}
The main difference  from using a CW pump source for the SFG nonlinear process is that the same heralding photon  can be coupled with multiple idler photons via using different pump waves that satisfy the energy conservation. Following the same approach outlined for the CW case, the spatial propagation of the heralding photon is
\begin{equation}
    \frac{\partial \hat{a}^{\dagger}_{h}}{\partial z}= -j\frac{\chi^{(2)}\Delta \omega_{i,h} }{2c}\sum_{i}A_{p}\sqrt{\frac{ \omega_{i}\omega_{h}}{n_{i}n_{h} } }\hat{a}^{\dagger}_{i}(z)e^{-j\kappa_{p}z} ,
    \label{eqn:Pulsedheraldcreation}
\end{equation}
where $A_{p}$ is the amplitude of the pump component with frequency $\omega_{p}=\omega_{h}-\omega_{i}$,  $\Delta \omega_{i,h}=\sqrt{\Delta \omega_{h}\Delta \omega_{i}}$, and  $\Delta \omega_{i}$ and $\Delta \omega_{h}$ the frequency sampling of the idler and the heralding photons, respectively.

\subsubsection{Total effects} 
Equation (\ref{eqn:SFWMaspulsed}) suggests that each signal photon is coupled  with all possible idler photons, which satisfy $\omega_{i}=\omega_{p_{1}}+\omega_{p_{2}}-\omega_{s}$. This would lead to a very large number of coupled differential equations. This case  represents the operation in the high-gain regime that can be approached under  strong pump excitation \cite{C6,quesada_effects_2014}. Similarly, Eq. (\ref{eqn:Pulsedheraldcreation}) suggests that  each heralding photon should be solved simultaneously with all possible idler photons, which satisfy $\omega_{h}=\omega_{i}+\omega_{p}$. However, practically only very few terms that satisfy the phase-matching condition would contribute to the dynamics of the SFG process. Assuming operating in the  low-gain regime and narrow-band phase matching condition, the system of coupled equations can be reduced to only three equations. In other words, one frequency triplet at a time ($\omega_{s},\omega_{i},\omega_{h}$) will be solved. In this case, the interaction between $\chi^{(2)}$ and $\chi^{(3)}$ in the case of a pulsed pump source can be approximated as

\begin{equation}
 \begin{split}
     &\frac{\partial \hat{a}_{s}}{\partial z}=j\kappa_{s}\hat{a}_{s}+j\sum_{p_{1}}\gamma_{s,i}\hat{a}_{i}^{\dagger}e^{j\left(\kappa_{p_{1}}+\kappa_{p_{2}}\right)z},\\
     &\frac{\partial \hat{a}_{i}^{\dagger}}{\partial z} =  -j\kappa_{i}\hat{a}_{i}^{\dagger} -j\sum_{p_{1}} \gamma_{s,i}\hat{a}_{s}e^{-j(\kappa_{p_{1}}+\kappa_{p_{2}})z} \\
   & \hphantom{\frac{\partial \hat{a}_{i}^{\dagger}}{\partial z} =}  -j\gamma_{i,h}\hat{a}_{h}^{\dagger}e^{j\kappa_{p_{3}}z},\\
     &   \frac{\partial \hat{a}_{h}^{\dagger}}{\partial z}=-j\kappa_{h}\hat{a}_{h}^{\dagger}-j\gamma_{i,h}\hat{a}_{i}^{\dagger}e^{-j\kappa_{p_{3}}z},
     \end{split}
     \label{eqn:PulsedQmodel}
 \end{equation}
where $\kappa_{p_{3}}$ is the propagation constant of the pump component involved in the SFG process with a frequency $\omega_{p_{3}}=\omega_{h}-\omega_{i}$, $\gamma_{s,i}=\frac{ 2 n_{2} \sqrt{n_{p_{1}}n_{p_{2}}  I_{p_{1}} I_{p_{2}}}\Delta \omega_{s,i}}{\Delta \omega_{p} \;c} \sqrt{\frac{\omega_{s} \omega_{i}}{n_{s} n_{i}}} $, and $\gamma_{i,h}=\frac{\chi^{(2)}A_{p}\Delta \omega_{i,h}}{2c}\sqrt{\frac{ \omega_{i}\omega_{h}}{n_{i}n_{h} }} $. In these formulas, we used the relation $3 \chi^{(3)} A_{p_1} A_{p_2}= 8 n_{2} \sqrt{n_{p_{1}}n_{p_{2}}  I_{p_{1}}I_{p_{2}}}$, where the pump intensity of a component with frequency $\omega_{p}$ is $I_p=\frac{Q \tau \Delta \omega^2}{2 \pi \sqrt{\pi} } e^{-\tau^2\left(\omega_p-\omega_{p_0}\right)^2}$, and $Q$ is the pulse energy. 
 
\subsubsection{Phase transformations} 
Applying the phase transformations  Eq. (\ref{eqn:SecondPhaseTransformation}) is not straightforward for a pulsed pump, because $\Delta k_{1}$ is not constant for different pump components in this case. We implement the following procedure to overcome this issue. In the summation $\sum_{p_{1}}$, only  terms with amplitudes larger than 0.15 of the maximum value of $\gamma_{s,i}$ are considered. These are the dominant terms in the summation. The phases $\Delta k_{1}$ associated with these terms are very close. Hence, these phases can be replaced by their average value, and the summation  $\sum_{p_{1}}$ can be approximated by a single phase term multiplied by a sum of the dominant amplitudes. We  examined this procedure by solving the approximated and exact differential equations  for different SFWM without involving SFG processes, and very similar results are always obtained using the two approaches. This  allows us to apply the phase transformations in Eq. (\ref{eqn:SecondPhaseTransformation}), and hence a similar set of coupled differential equations Eq. (\ref{eqn:QuamtumModel}) is  attained. 

\subsubsection{Expected number of photons}
Since the definitions of the coupling coefficients  $\gamma_{s,i}$ and $\gamma_{i,h}$ include $\Delta \omega_{s,i}$ and $\Delta \omega_{i,h}$ , the expected number of photons is dimensionless and  represent the mean number of photons in a pixel with an area of $\Delta \omega_{s} \Delta \omega_{i}$ for $ \langle N_{s}(L)\rangle$ and $ \langle N_{i}(L)\rangle$, and a pixel with an area of $\Delta \omega_{i} \Delta \omega_{h}$ for $ \langle N_{h}(L)\rangle$. It is worth  noting that the joint spectral intensity (JSI) is related to the expected number of photons via $ \langle N_{q}(L)\rangle/\Delta \omega_{q}^2$. Also, the off-diagonal elements of the matrix $\textbf{T}$ can be used to calculate the joint spectral amplitude (JSA) and the spectral purity of the photon pairs \cite{C6,C7,C8}. The expected number of photons or JSI will reside in a 3D space with $\omega_{s}$, $\omega_{i}$, and $\omega_{h}$ being its bases \cite{C16}. Since our main focus is on the JSI of the signal and heralding photons, the JSI  can be integrated across the idler frequency spectrum, results in a 2D function. The outcome of this step also discloses the marginal probability of obtaining a heralding photon at a certain frequency regardless of the idler frequency \cite{ndagano_imaging_2020,lange_cryogenic_2022}. 

In the event when all the idler photons are successfully up-converted, the expected counts for both signal and heralding photons should match,  each reflecting the same information about the signal-heralding pairs. In this case, either count can be used to assess the purity of these pairs. However, when partial up-conversion takes place the purity of these pairs is determined using the expected heralding photon count, since the expected signal photon count includes information about  heralding photons and also idler photons that are not up-converted.

% knowing this will be helpful in calculating the spectral purity of the photon pairs.

\section{Simulation results}
In this section, we focus on the design of waveguides that facilitate the study of the interaction between the SFWM and SFG. This interplay is examined  in AlGaAs and LiNbO$_3$ nanophotonic waveguides. We begin by solely investigating the SFWM in these waveguides via calculating the expected number of photons across various coupled frequencies and assessing the purity of the generated photon pairs under different conditions. Subsequently, we delve into enhancing the SFG process within the waveguide by applying  quasi-phase matching techniques such as periodic poling in LiNbO$_3$ or orientation pattern in AlGaAs. This  allows the idler photons to act as an intermediate stage that combines  the  SFWM and SFG processes via being  annihilated and  up-converted to the heralding photons with higher frequencies for efficient detection. We also calculate the purity of the signal-heralding photon pairs, providing insights into the interplay and efficiency of these two simultaneous nonlinear optical processes within the designed waveguide.

\subsection{AlGaAs buried waveguide}

 AlGaAs is a non-centrosymmetric material with strong $\chi^{(2)}$ and $\chi^{(3)}$ nonlinear coefficients, essential for supporting the simultaneous occurrence of SFWM and SFG. An $\mathrm{Al_{0.3}Ga_{0.7}As}$ rectangular waveguide with  300 nm thickness, 600 nm width,  buried in silica cladding, $\chi^{(2)}=120\times10^{-12}$m/V \cite{C9,ohashi_determination_1993}, and $n_2=10^{-17} $m$^2$/W \cite{apiratikul_enhanced_2014} is assumed in the simulations of this section, as shown in Fig. \ref{fig1}(a).

\subsubsection{Spontaneous Four-Wave Mixing}
\begin{figure}[H]
\centering
\includegraphics[width=\columnwidth]{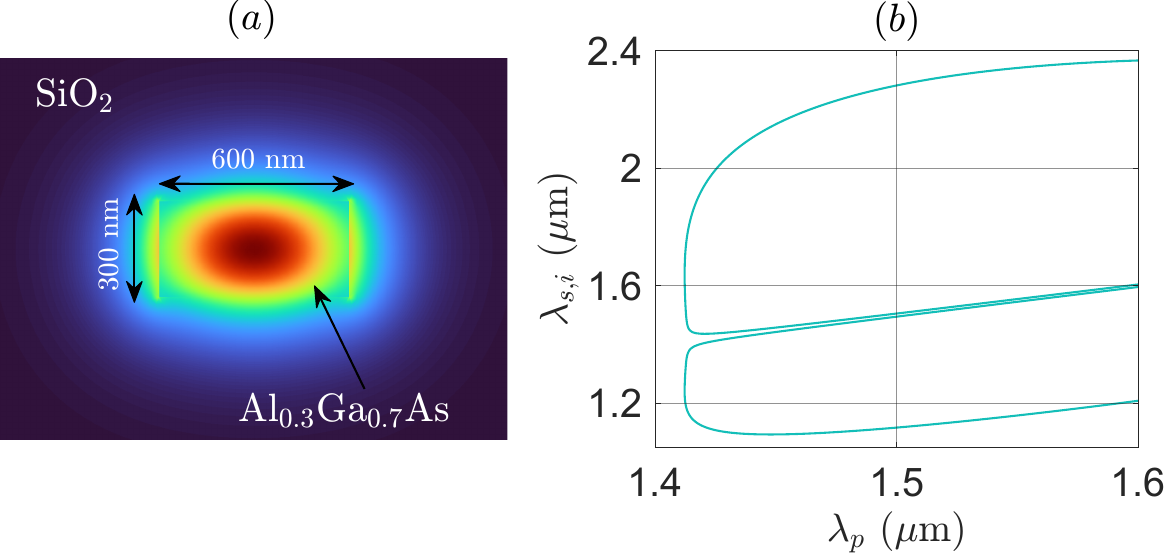}
\caption[fig1]{ (a) Cross-sectional area of the AlGaAs waveguide with the fundamental TE mode at 1550 nm wavelength. (b) The phase-matching contour of a SFWM process $2\omega_{p}=\omega_{s}+\omega_{i}$, assuming a monochromatic pump source of power 100 mW.
% where the pump was assumed to be monochromatic with a power of 100 mW. 
}
\label{fig1}
\end{figure}

We assume that the pump, signal and idler are all in the TE fundamental mode for a SFWM interaction. The effective refractive indices of these modes that account for both the material and the waveguide dispersion have been calculated using COMSOL’s mode solver. The efficiency of the SFWM is mainly influenced by the phase-matching condition and the pump power. Figure \ref{fig1}(b) depicts the phase-matching contour for the SFWM process. At a certain pump wavelength there are two possible distinct signal and idler pairs,  close to and away from the pump photons.

\begin{figure}[H]
    \includegraphics[width=\columnwidth]{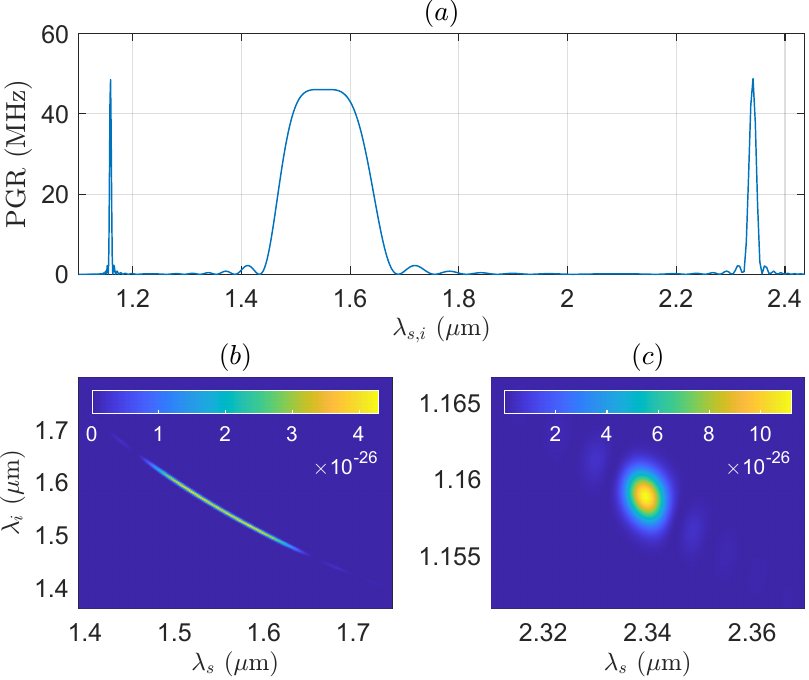} 
    \caption{(a) Expected photon count or photon generation rate (PGR) of the generated pairs in the AlGaAs waveguide  with length 1.7 mm. (b)  Joint spectral intensity of the photon pairs generated close  to the pump central wavelength in a waveguide with length 3 mm. (c)  Joint spectral intensity of the photon pairs generated far from the pump central wavelength in a waveguide with length 4.5 mm. The colorbar denotes the probability density in units Hz$^{-2}$.}
    \label{fig:2} 
\end{figure}

The expected number of photons generated via considering only the SFWM process, i.e. $\langle N_s\rangle=\langle N_i\rangle$, at the waveguide output is displayed in Fig. \ref{fig:2}(a) for  a CW pump source with a wavelength 1550 nm and power 100 mW.  In this case, $\langle N_s\rangle$  denotes the average number of photon pairs produced per unit time per unit frequency. To compare the rate of pair generation calculated in the simulation with the experimental results, $\langle N_s\rangle$ is multiplied by a filter spectral resolution of a typical value of 200 GHz \cite{C13,C14}. Following the phase-matching  contour in Fig. \ref{fig1}, two distinct signal and idler pairs are observed, as depicted in Fig. \ref{fig:2}(a). The first pair is far from the pump, exhibits a narrow bandwidth and is centered around  signal and idler wavelengths 2341 nm and 1159 nm, respectively.   The second pair is centered around the pump wavelength and spans over a broad bandwidth.

The joint spectral intensity (JSI) for a pulsed source with a Gaussian profile, energy $1.064$ pJ, central wavelength 1550 nm, and a temporal width 0.6 ps is portrayed in Fig. \ref{fig:2}(b,c) for the close and far photon-pairs. Different waveguide lengths have been used for the two cases. Applying the singular-value decomposition technique to the elements of the matrix $\textbf{T}_{12}\left(\omega_s,\omega_i\right)$, the spectral purity of the generated photon pairs can be calculated \cite{C15}. The spectral purity for the JSI shown in Fig. \ref{fig:2}(b), assuming the signal and idler wavelengths are filtered at 1559 nm and 1541 nm, is 0.0455. The reason for the low purity is the orientation of the phase matching function at these particular wavelengths. On the other hand, the spectral purity for the far-peak case is 0.87, which is  high  but  not optimal due to the presence of the sinc function tails. These tails can be eliminated and hence the spectral purity can be enhanced by spectral filtering, for instance. 

\subsubsection{Interplay between SFWM and SFG}
\begin{figure}[H]
\centering
    \includegraphics[width=0.95\columnwidth]{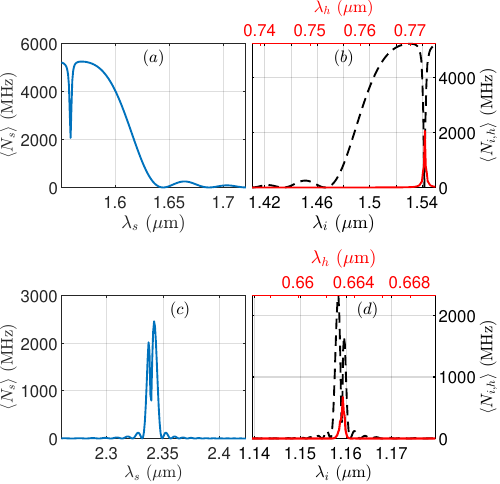} 
    \caption{Expected signal (blue line), idler (dashed line) and heralding (red line) photon rate in a 3.5 mm long AlGaAs waveguide pumped with a CW source at a wavelength 1550 nm, power 300 mW, and poling period 3.53 \textmu m (a,b), 2.94 \textmu m (c,d).
     }
    \label{fig:3} 
\end{figure}

\begin{figure} [H]
\centering
    \includegraphics[width=0.9\columnwidth]{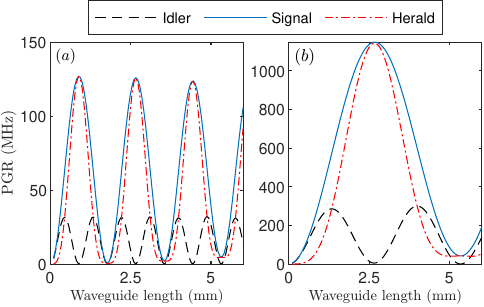} 
    \caption{The spatial dependence of the expected photon rates generated by the interplay between the SFWM and SFG in the AlGaAs waveguide pumped by a CW source at 1550 nm with a power of 0.5 W  using a modulation period of (a)  1.18 \textmu m and (b) 3.53 \textmu m.} 
    \label{fig:L} 
\end{figure}

To enhance the SFG process between the idler and the pump photons, the second-order $\chi^{(2)}$ nonlinear coefficient can be periodically modulated via the orientation-pattern technique to allow for a quasi-phase matching between the interacting photons. The up-converted (or heralding) photon is assumed to be in the fundamental TM mode, since the $\chi^{(2)} $ tensor properties of AlGaAs do not allow all interacting photons to have the same polarisation \cite{C9}. The expected signal, idler, and heralding  photon rates under a CW-pump excitation with a wavelength 1550 and  power  300 mW along a waveguide with a modulation period 3.53 \textmu m, which is the third-order harmonic of the fundamental period $2\pi/\Delta\kappa_2$, is shown in Fig. \ref{fig:3}(a,b). The period is chosen to up-convert the idler photons generated close to the pump. As portrayed, the rate of the heralding photons at 772 nm is 99\% of the signal photons at 1558.6 nm, which indicates that approximately 99\% of the idler photon at 1541 nm is up-converted to heralding photons. Similar behavior is obtained in the case of up-converting the idler photon centered at 1159 nm as displayed in Fig. \ref{fig:3}(c,d) using a modulation period 2.94 \textmu m. We find that it is important to choose an appropriate order of the modulation period to balance the interplay between the SFWM and SFG, otherwise the $\chi^{(2)} $ process would be dominant. In the latter case, the idler and heralding photons will periodically exchange energy over a short distance, preventing the heralding photons from exponentially growing to higher values, as portrayed in Fig. \ref{fig:L}.

\begin{figure}[H]
\centering
    \includegraphics{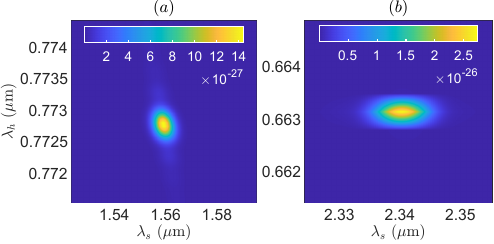} 
    \caption{ Expected number of heralding photons $\langle N_{h} \rangle$  generated via up-converting the idler photons: centered at $1541$ nm using a modulation period 3.53 \textmu m (a),  centered at $1159$ nm using a modulation period 2.94 \textmu m (b). The colorbar denotes the probability density in units Hz$^{-2}$.}
    \label{fig:4} 
\end{figure}

The expected number of the heralding photons $ \langle N_{h}(L)\rangle $ for different combinations of $\omega_s$ and $\omega_h$ or the 2D JSI is displayed in Fig. \ref{fig:4}. Panels (a,b) represent up-converting the idler photons at wavelength 1541 nm and 1159 nm employing the pulse and waveguide parameters used in Figs. \ref{fig:2}(b,c), respectively. The associated purity of the former case is interesting high 0.9, in contrast to the significantly low purity 0.0455 obtained directly via SFWM without involving the SFG process. Similarly, the purity of the latter case also is improved from 0.87 to 0.99, indicating the advantage of implementing the up-conversion technique as a novel technique in enhancing the single-photon spectral purity.

\subsection{LiNbO$_{3}$ ridge waveguide}
\begin{figure}[H]
\centering
    \includegraphics[width=\columnwidth]{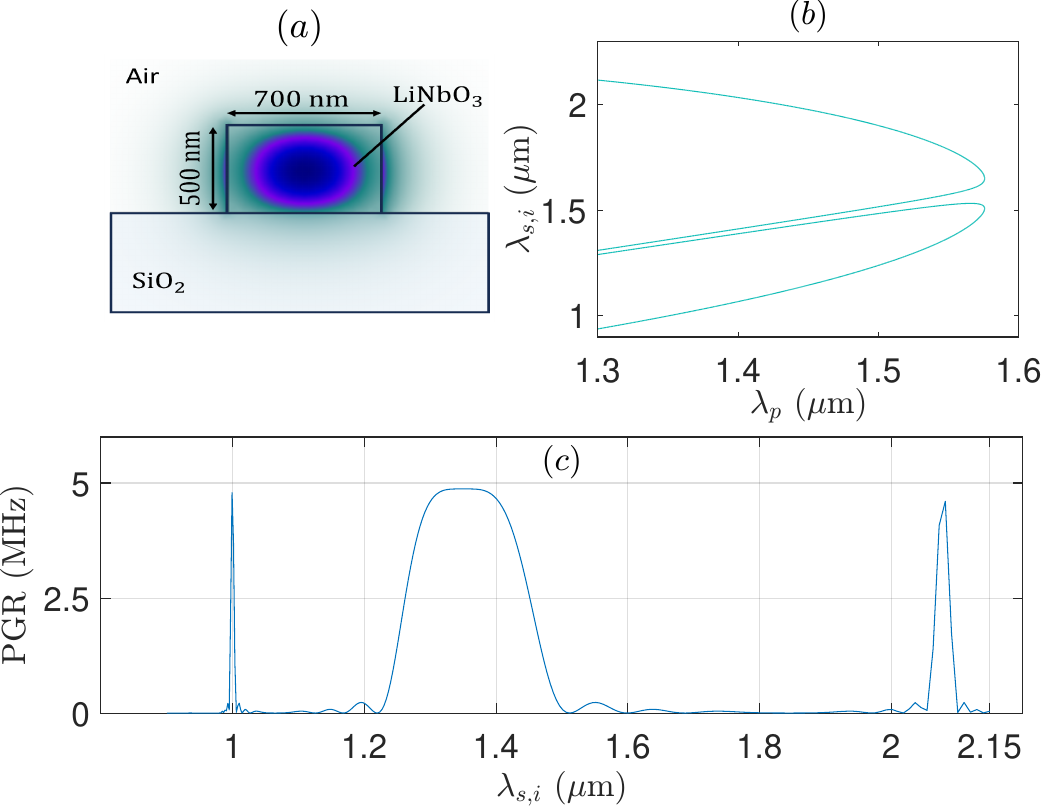} 
    \caption{ (a) A cross section of a lithium-niobate waveguide used in the simulations with the fundamental TE mode at 1350 nm enclosed. (b) Phase-matching contour of the SFWM process with the pump, signal and idler in the fundamental TE mode. (c) Expected number of photons generated in a 1 mm-long waveguide  pumped by a CW source at a wavelength 1350 nm.}
    \label{fig:5} 
\end{figure}

Lithium niobate,  LiNbO$_{3}$, is another potential platform that can allow for a concurrent interplay between SFWM and SFG paramertic processes. Lithium niobate is an attractive photonic material because of its broadband transparency window from approximately 0.4 \textmu m to 4.5 \textmu m, large electro- and nonlinear-optic coefficients, and compatibility with the electric-field poling technique to obtain quasi-phase matching between different waves via  periodically flipping the $\chi^{(2)}-$nonlinear coefficient \cite{myers_quasi-phase-matched_1995}. We  apply our model to a LiNbO$_{3}$ ridge waveguide, as shown in Fig. \ref{fig:5}(a), with a 500 nm thickness, a 700 nm width, a silica substrate, $\chi^{(2)}=25\times10^{-12}$m/V \cite{C9}, and $n_2=2.6\times 10^{-19} $m$^2$/W \cite{jankowski_ultrabroadband_2020}. Unlike AlGaAs, the tensor properties of $\chi^{(2)}$ in LiNbO$_{3}$ can allow its strongest component to mediate a second-order nonlinear interaction between photons with the same polarisation. Hence, we choose all interacting photons (pump, signal, idler and heralding) to be in the same fundamental TE mode. The phase-matching contour for the SFWM is depicted in Fig. \ref{fig:5}(b) showing, similar to the AlGaAs waveguide, the possibility of generating two distinct signal and idler photons  close and far from the pump central wavelength. The expected number of photons generated by the SFWM using a CW pump with a wavelength 1350 nm and a power 5 W is portrayed in Fig. \ref{fig:5}(c). The pairs generated far from the pump wavelength are centered at signal and idler wavelengths of 2082 nm and 999 nm, respectively.

% \begin{figure*}
%     \includegraphics[width=0.92\textwidth]{graphics/fig5-new.pdf} 
%     \caption{ (a) Cross section of the simulated waveguide. (b) The phase-matching contour for the SFWM process with the pump, signal and idler functioning in the fundamental TE mode. (c) Expected number of photons generated at the output of a 1 mm waveguide pumped by a cw source at a wavelength of 1350 nm.}
%     \label{fig:5} 
% \end{figure*}

\begin{figure}[H]
    \centering 
    \includegraphics[width=\columnwidth]{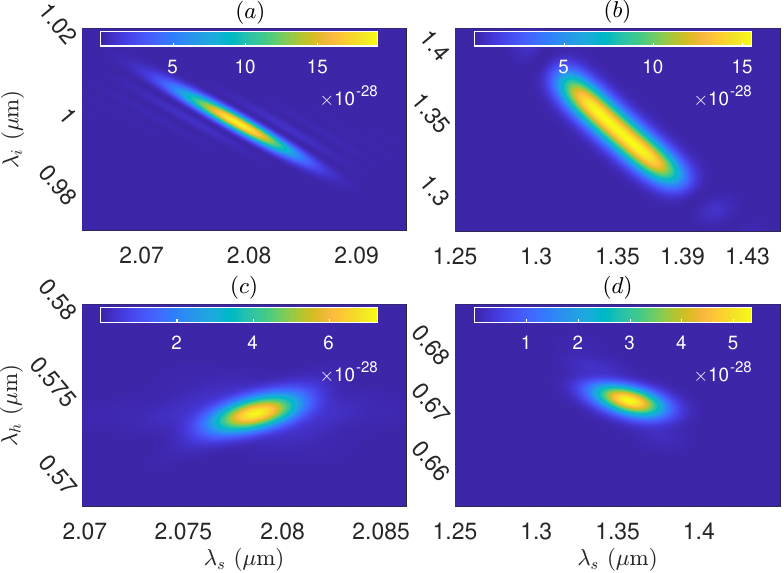} 
    \caption{Expected  number of signal-idler photon pairs via the SFWM process at wavelengths far (a) and near  (b) from the pump central wavelength.  Expected  number of signal-heralding photon pairs via the SFWM and SFG processes at wavelengths far (c) and near  (d) from the pump central wavelength. The colorbar denotes probability density in Hz$^{-2}$.}
    \label{fig:6} 
\end{figure}

For pulse sources in the absence of SFG, the expected number of generated signal and idler pairs via SFWM, $\langle N_s\rangle=\langle N_i\rangle$, is displayed  in Fig. \ref{fig:6}(a,b). Panel (a) shows the case of the photon pairs generated away from the pump central wavelength using a 10 pJ pump pulse centered at 1350 nm and with a temporal width 114 fs in a 4 mm-long waveguide. The expected number of another photon-pairs  selected at 1344 nm and 1356 nm (close to  the pump central wavelength 1350 nm) is depicted in \ref{fig:6}(b) using  a 4.5 mm-long waveguide and a pump pulse  with an energy 8 pJ and temporal duration 91 fs. The spectral purities for the both cases in Figs. \ref{fig:6}(a,b) is 0.2 and 0.45, respectively, which is very low. However, allowing the up-conversion of the idler photons to the heralding photons using poling periods of $\Lambda=5.82 $ \textmu m and $\Lambda=5.95$ \textmu m, significantly enhance the spectral purities to  0.94 and 0.86, as illustrated in Figs. \ref{fig:6}(c,d).

In comparison to the AlGaAs waveguide the expected number of photons in the LiNbO$_3$ waveguide is lower, since the former material has higher both second and third-order nonlinearities. However, the requirement of having a modulation period in the few-micrometer range in AlGaAs based on the orientation-pattern fabrication technique might be currently a hurdle, which is not the case using the electric-field poling technique. But we envisage this range of periods can be  attainable soon in semiconductor materials, as the fabrication techniques rapidly evolve.

\section{Conclusion}

In this work, we develop a robust quantum model to study the interplay between the SFWM and SFG processes in nanophotonic  waveguides with strong quadratic and cubic nonlinearities. We exploit this interplay in developing on-chip pure heralded single photon sources that can be easily complemented with, for instance, on-chip silicon photon detectors for the heralding process. Moreover, we show how the up-conversion process significantly enhances the spectral purity of the signal-idler pairs generated by the SFWM process. The provided theoretical models and analyses lay the groundwork for future experimental validations and technological developments. The findings offer new avenues for research in optical quantum computing and quantum secure communications applications. We examine our proposed device using AlGaAs and LiNbO$_3$ nanophotonic waveguides, however, further exploration into waveguide materials and geometries will help in realising the full potential of this introduced approach.

\begin{backmatter}
\bmsection{Funding} EPSRC Doctoral Training Partnership (DTP), and Institute of Photonics and Quantum Sciences at Heriot-Watt University.

\bmsection{Disclosures} Authors state no conflict of interest.

\bmsection{Data availability} Data underlying the results presented in this paper are not publicly available at this time but may be obtained from the authors upon reasonable request.

\end{backmatter}

\bibliography{references}

\begin{thebibliography}{10}
\newcommand{\enquote}[1]{``#1''}

\bibitem{c01}
E.~Pelucchi \emph{et~al.}, \enquote{The potential and global outlook of integrated photonics for quantum technologies,} {\protect\JournalTitle{Nature Reviews Physics}} \textbf{4}, 194--208 (2022). Number: 3 Publisher: Nature Publishing Group.

\bibitem{c02}
J.~Wang, F.~Sciarrino, A.~Laing, and M.~G. Thompson, \enquote{Integrated photonic quantum technologies,} {\protect\JournalTitle{Nature Photonics}} \textbf{14}, 273--284 (2020). Number: 5 Publisher: Nature Publishing Group.

\bibitem{c03}
J.~W. Silverstone, D.~Bonneau, J.~L. O’Brien, and M.~G. Thompson, \enquote{Silicon {Quantum} {Photonics},} {\protect\JournalTitle{IEEE JOURNAL OF SELECTED TOPICS IN QUANTUM ELECTRONICS}} \textbf{22} (2016).

\bibitem{c04}
P.~Lodahl, \enquote{Quantum-dot based photonic quantum networks,} {\protect\JournalTitle{Quantum Science and Technology}} \textbf{3}, 013001 (2017). Publisher: IOP Publishing.

\bibitem{PhysRevLett.132.133603}
H.~Zeng, Z.-Q. He, Y.-R. Fan, \emph{et~al.}, \enquote{Quantum light generation based on gan microring toward fully on-chip source,} {\protect\JournalTitle{Phys. Rev. Lett.}} \textbf{132}, 133603 (2024).

\bibitem{PhysRevLett.133.083803}
R.~Chen, Y.-H. Luo, J.~Long, \emph{et~al.}, \enquote{Ultralow-loss integrated photonics enables bright, narrowband, photon-pair sources,} {\protect\JournalTitle{Phys. Rev. Lett.}} \textbf{133}, 083803 (2024).

\bibitem{c05}
I.~Aharonovich, D.~Englund, and M.~Toth, \enquote{Solid-state single-photon emitters,} {\protect\JournalTitle{Nature Photonics}} \textbf{10}, 631--641 (2016). Number: 10 Publisher: Nature Publishing Group.

\bibitem{c07}
P.~Senellart, G.~Solomon, and A.~White, \enquote{High-performance semiconductor quantum-dot single-photon sources,} {\protect\JournalTitle{Nature Nanotechnology}} \textbf{12}, 1026--1039 (2017). Number: 11 Publisher: Nature Publishing Group.

\bibitem{c06}
M.~D. Eisaman, J.~Fan, A.~Migdall, and S.~V. Polyakov, \enquote{Invited review article: Single-photon sources and detectors,} {\protect\JournalTitle{Review of Scientific Instruments}} \textbf{82}, 071101 (2011).

\bibitem{c08}
C.~Joshi, A.~Farsi, S.~Clemmen, \emph{et~al.}, \enquote{Frequency multiplexing for quasi-deterministic heralded single-photon sources,} {\protect\JournalTitle{Nature Communications}} \textbf{9}, 847 (2018). Number: 1 Publisher: Nature Publishing Group.

\bibitem{c09}
E.~Meyer-Scott, C.~Silberhorn, and A.~Migdall, \enquote{Single-photon sources: {Approaching} the ideal through multiplexing,} {\protect\JournalTitle{Review of Scientific Instruments}} \textbf{91}, 041101 (2020).

\bibitem{Yu:22}
H.~Yu, C.~Yuan, R.~Zhang, \emph{et~al.}, \enquote{Spectrally multiplexed indistinguishable single-photon generation at telecom-band,} {\protect\JournalTitle{Photon. Res.}} \textbf{10}, 1417--1429 (2022).

\bibitem{apiratikul_enhanced_2014}
P.~Apiratikul \emph{et~al.}, \enquote{Enhanced continuous-wave four-wave mixing efficiency in nonlinear {AlGaAs} waveguides,} {\protect\JournalTitle{Optics Express}} \textbf{22}, 26814--26824 (2014). Publisher: Optica Publishing Group.

\bibitem{c014}
H.~Mahmudlu, S.~May, A.~Angulo, \emph{et~al.}, \enquote{{AlGaAs}-on-insulator waveguide for highly efficient photon-pair generation via spontaneous four-wave mixing,} {\protect\JournalTitle{Optics Letters}} \textbf{46}, 1061--1064 (2021). Publisher: Optica Publishing Group.

\bibitem{c013}
S.~Lauria and M.~F. Saleh, \enquote{Mixing second- and third-order nonlinear interactions in nanophotonic lithium-niobate waveguides,} {\protect\JournalTitle{Physical Review A}} \textbf{105}, 043511 (2022). Publisher: American Physical Society.

\bibitem{thomas_quasi_2013}
J.~Thomas, V.~Hilbert, R.~Geiss, \emph{et~al.}, \enquote{Quasi phase matching in femtosecond pulse volume structured x-cut lithium niobate,} {\protect\JournalTitle{Laser \& Photonics Reviews}} \textbf{7}, L17--L20 (2013). \_eprint: https://onlinelibrary.wiley.com/doi/pdf/10.1002/lpor.201200116.

\bibitem{morais_directionally_2017}
N.~Morais \emph{et~al.}, \enquote{Directionally induced quasi-phase matching in homogeneous {AlGaAs} waveguides,} {\protect\JournalTitle{Optics Letters}} \textbf{42}, 4287--4290 (2017). Publisher: Optica Publishing Group.

\bibitem{zhang_high_performance_2021}
Z.~Zhang, C.~Yuan, S.~Shen, \emph{et~al.}, \enquote{High-performance quantum entanglement generation via cascaded second-order nonlinear processes,} {\protect\JournalTitle{npj Quantum Information}} \textbf{7}, 1--9 (2021). Publisher: Nature Publishing Group.

\bibitem{c015}
S.~Gyger \emph{et~al.}, \enquote{Reconfigurable photonics with on-chip single-photon detectors,} {\protect\JournalTitle{Nature Communications}} \textbf{12}, 1408 (2021). Number: 1 Publisher: Nature Publishing Group.

\bibitem{c016}
W.~Luo \emph{et~al.}, \enquote{Recent progress in quantum photonic chips for quantum communication and internet,} {\protect\JournalTitle{Light: Science \& Applications}} \textbf{12}, 175 (2023). Number: 1 Publisher: Nature Publishing Group.

\bibitem{c017}
M.~Davanco \emph{et~al.}, \enquote{Heterogeneous integration for on-chip quantum photonic circuits with single quantum dot devices,} {\protect\JournalTitle{Nature Communications}} \textbf{8}, 889 (2017). Number: 1 Publisher: Nature Publishing Group.

\bibitem{c018}
G.~Moody \emph{et~al.}, \enquote{2022 {Roadmap} on integrated quantum photonics,} {\protect\JournalTitle{Journal of Physics: Photonics}} \textbf{4}, 012501 (2022). Publisher: IOP Publishing.

\bibitem{C1}
B.~Huttner, S.~Serulnik, and Y.~Ben-Aryeh, \enquote{Quantum analysis of light propagation in a parametric amplifier,} {\protect\JournalTitle{Phys. Rev. A}} \textbf{42}, 5594--5600 (1990).

\bibitem{C2}
M.~F. Saleh, \enquote{Modelling spontaneous four-wave mixing in periodically tapered waveguides,} {\protect\JournalTitle{Opt. Express}} \textbf{27}, 11979--11990 (2019).

\bibitem{C3}
K.~J. Blow, R.~Loudon, S.~J.~D. Phoenix, and T.~J. Shepherd, \enquote{Continuum fields in quantum optics,} {\protect\JournalTitle{Phys. Rev. A}} \textbf{42}, 4102--4114 (1990).

\bibitem{C4}
Y.~R. Shen, \enquote{Quantum statistics of nonlinear optics,} {\protect\JournalTitle{Phys. Rev.}} \textbf{155}, 921--931 (1967).

\bibitem{C6}
G.~Triginer \emph{et~al.}, \enquote{Understanding high-gain twin-beam sources using cascaded stimulated emission,} {\protect\JournalTitle{Phys. Rev. X}} \textbf{10}, 031063 (2020).

\bibitem{quesada_effects_2014}
N.~Quesada and J.~E. Sipe, \enquote{Effects of time ordering in quantum nonlinear optics,} {\protect\JournalTitle{Physical Review A}} \textbf{90}, 063840 (2014). Publisher: American Physical Society.

\bibitem{C7}
O.~Alibart, J.~Fulconis, G.~K.~L. Wong, \emph{et~al.}, \enquote{Photon pair generation using four-wave mixing in a microstructured fibre: theory versus experiment,} {\protect\JournalTitle{New Journal of Physics}} \textbf{8}, 67 (2006).

\bibitem{C8}
K.~Garay-Palmett, A.~B. U'Ren, and R.~Rangel-Rojo, \enquote{Conversion efficiency in the process of copolarized spontaneous four-wave mixing,} {\protect\JournalTitle{Phys. Rev. A}} \textbf{82}, 043809 (2010).

\bibitem{C16}
D.~L.~P. Vitullo, M.~G. Raymer, B.~J. Smith, \emph{et~al.}, \enquote{Entanglement swapping for generation of heralded time-frequency-entangled photon pairs,} {\protect\JournalTitle{Phys. Rev. A}} \textbf{98}, 023836 (2018).

\bibitem{ndagano_imaging_2020}
B.~Ndagano \emph{et~al.}, \enquote{Imaging and certifying high-dimensional entanglement with a single-photon avalanche diode camera,} {\protect\JournalTitle{npj Quantum Information}} \textbf{6}, 1--8 (2020). Publisher: Nature Publishing Group.

\bibitem{lange_cryogenic_2022}
N.~A. Lange \emph{et~al.}, \enquote{Cryogenic integrated spontaneous parametric down-conversion,} {\protect\JournalTitle{Optica}} \textbf{9}, 108--111 (2022). Publisher: Optica Publishing Group.

\bibitem{C9}
B.~Saleh and M.~Teich, \emph{Fundamentals of Photonics, 2nd Edition} (2007).

\bibitem{ohashi_determination_1993}
M.~Ohashi \emph{et~al.}, \enquote{Determination of quadratic nonlinear optical coefficient of {AlxGa1}−{xAs} system by the method of reflected second harmonics,} {\protect\JournalTitle{Journal of Applied Physics}} \textbf{74}, 596--601 (1993).

\bibitem{C13}
C.~Xiong, L.~G. Helt, A.~C. Judge, \emph{et~al.}, \enquote{Quantum-correlated photon pair generation in chalcogenide as2s3 waveguides,} {\protect\JournalTitle{Opt. Express}} \textbf{18}, 16206--16216 (2010).

\bibitem{C14}
K.~Park, D.~Lee, R.~W. Boyd, and H.~Shin, \enquote{Telecom c-band photon-pair generation using standard smf-28 fiber,} {\protect\JournalTitle{Optics Communications}} \textbf{484}, 126692 (2021).

\bibitem{C15}
P.~J. Mosley, J.~S. Lundeen, B.~J. Smith, and I.~A. Walmsley, \enquote{Conditional preparation of single photons using parametric downconversion: a recipe for purity,} {\protect\JournalTitle{New Journal of Physics}} \textbf{10}, 093011 (2008).

\bibitem{myers_quasi-phase-matched_1995}
L.~E. Myers \emph{et~al.}, \enquote{Quasi-phase-matched optical parametric oscillators in bulk periodically poled {LiNbO}$_{\textrm{3}}$,} {\protect\JournalTitle{JOSA B}} \textbf{12}, 2102--2116 (1995). Publisher: Optica Publishing Group.

\bibitem{jankowski_ultrabroadband_2020}
M.~Jankowski \emph{et~al.}, \enquote{Ultrabroadband nonlinear optics in nanophotonic periodically poled lithium niobate waveguides,} {\protect\JournalTitle{Optica}} \textbf{7}, 40--46 (2020). Publisher: Optica Publishing Group.

\end{thebibliography}

\end{document}